\documentclass{INTERSPEECH2023}

\interspeechcameraready

\title{Implementing contextual biasing in GPU decoder for online ASR}
\name{Iuliia Nigmatulina$^{1,2}$, Srikanth Madikeri$^{1}$, Esaú Villatoro-Tello$^{1}$, Petr Motlicek$^{1,3}$ \\Juan Zuluaga-Gomez$^{1,4}$, Karthik Pandia$^5$, Aravind Ganapathiraju$^5$\thanks{This work was supported by the Idiap\&Uniphore collaboration project and partially by CRiTERIA (EC Horizon 2020, n.: 101021866) and ROXANNE (EC Horizon 2020, n.: 833635).}}

\address{
  $^1$Idiap Research Institute, Switzerland $^2$University of Zurich, Switzerland \\
  $^3$ Faculty of Information Technology, Brno University of Technology, Czech Republic \\
  $^4$LIDIAP, Ecole Polytechnique Federale de Lausanne (EPFL), Switzerland \\
  $^5$Uniphore, India
  }

\email{iuliia.nigmatulina@idiap.ch}

\begin{document}

\maketitle
 
\begin{abstract}
GPU decoding significantly accelerates the output of ASR predictions. While GPUs are already being used for online ASR decoding, post-processing and rescoring on GPUs have not been properly investigated yet. Rescoring with available contextual information can considerably improve ASR predictions. Previous studies have proven the viability of lattice rescoring in decoding and biasing language model (LM) weights in offline and online CPU scenarios. In real-time GPU decoding, partial recognition hypotheses are produced without lattice generation, which makes the implementation of biasing more complex. The paper proposes and describes an approach to integrate contextual biasing in real-time GPU decoding while exploiting the standard Kaldi GPU decoder. Besides the biasing of partial ASR predictions, our approach also permits dynamic context switching allowing a flexible rescoring per each speech segment directly on GPU. The code is publicly released\footnote{\url{https://github.com/idiap/contextual-biasing-on-gpus}} and tested with open-sourced test sets.
\end{abstract}
\noindent\textbf{Index Terms}: real-time speech recognition, contextual adaptation, GPU decoding, finite-state transducers

\section{Introduction}

Contextual biasing of ASR has proven to be useful for many applications where prior information is available. Typically, contextual biasing in ASR works by adjusting weights of the model, or costs of words in recognition lattices, and it has been used to improve recognition of named entities (NEs), such as contacts, locations, film titles, etc. \cite{aleksic2015bringing,hall2015composition,williams2017rescoring,serrino2019contextual,shore2012knowledge,oualil2017context,kocour21_interspeech,Iuliia_SUBMITTEDTOINTERSPEECH2021_2021,zuluagagomez21_interspeech,motlicek2010english}.

Real-time decoding can work on central processing units (CPUs), as well as on graphics processing units (GPUs) that can considerably accelerate decoding~\cite{braun2020gpu}. Although the previous biasing methods perform well, most experiments were done for decoding on CPUs and thus are missing possible acceleration of real-time transcribing available while decoding with GPUs. This paper focuses on (1)~contextual biasing in real-time decoding, (2)~dynamic integration of contextual information in online GPU-based decoders, and (3)~analysis of the relevance of parallelized contextual biasing.

For all experiments presented in the paper, we use Kaldi online decoders~\cite{braun2020gpu,povey2011kaldi}. In these decoders, lattice generation to pick the best output path is performed on the CPUs. Generation of real-time partial transcriptions works differently to avoid the generation of lattices, which
%is typically done on CPUs that
would considerably increase the latency to obtain the final system output. Thus, a GPU-based online decoder outputs partial predictions directly selecting the current best sequence of tokens, and allows the process to be fully parallelized. This implementation, however, makes it impossible to integrate standard methods of contextual biasing directly into the online GPU decoder, as they involve lattice rescoring with FST composition. Although the implementation of lattice composition on GPUs has been proposed before \cite{argueta2018composing,li2021parallelizable}, the main challenges of the current work is to find a rescoring method without directly dealing with lattices and avoiding lattice generation for partial hypotheses.

%\begin{figure}[t]
%  \centering
%  \includegraphics[width=\linewidth]{pictures/fst_pic.jpg}
%  \caption{A toy-example of biased FST with biasing sequences: \textbf{`ryanair one romeo kilo'} and \textbf{`turkish six one heavy'}.}
%  \label{fig:pic_fst}
%\end{figure}

Another challenge arises during the use of multiple contextual biases in the same decoder, where each utterance has its own biasing FST. Such a situation may arise, for instance, when another modality is providing the constantly changing contextual information for decoding\footnote{For example, the change of time, location, topic of conversation, etc. In the air traffic communication (ATC) domain, such information may come from radar data.}. It becomes prohibitively large to maintain multiple FSTs in-memory due to the nature of the composition even though the actual contextual information encoded amounts to few textual tokens. We address this challenge by keeping the contextual information independent of the original decoding graph by only storing a list of indices of the arcs to be boosted for each context. A discounting offset is added to these arcs only during decoding, thus indirectly composing the context graph with the decoding graph.

%where the question of leveraging contextual data to boost target entities (words and/or word sequences) is addressed.
%callsigns\footnote{Callsigns are unique identifiers for air crafts, of which the first part is an abbreviation of airline name and a last part is a flight number that contains a digit combination and may also incorporate an additional character combination, e.g., \textit{swiss two six eight nine}, \textit{ryanair one sierra golf}.} in ATC conversations for offline decoding.
Contextual information is usually passed as a small weighted finite state transducer (WFST) created from a list of words and/or word sequences to be boosted.
%(see example in Fig.~\ref{fig:pic_fst}).
In this paper, we extend the previous work on contextual biasing \cite{aleksic2015bringing,hall2015composition,williams2017rescoring,kocour21_interspeech,Iuliia_SUBMITTEDTOINTERSPEECH2021_2021,nigmatulina2022two}.
We investigate the suitability of existing biasing methods from the previous work for the GPU online decoder and propose a new approach to dynamically boost contextual information in the parallelized GPU decoding that does not require lattice generation. Our main contributions are (1)~ an analysis of possible ways to use contextual information in online parallelized decoding, and (2)~the first publicly available implementation of dynamic contextual biasing in an online GPU decoder.

%The rest of the paper is organized as follows: Section 2 reviews current GPU decoders and the main implementation features of the Kaldi-Nvidia GPU decoder. Section 3 gives a theoretical background of existing biasing methods, and explains the motivation behind the proposed biasing method for distributed decoding. Section 4 describes the implementation details of the proposed method. Then, we present the data and the experiment set up in Section 5. We report the results, analyze our observations, and discuss our ideas in Sections 6 and 7, respectively. 

\section{Online decoding on CPUs vs GPUs}
\subsection{Decoding on CPUs}
Online decoding on CPUs (for example, Kaldi's online2-tcp-nnet3-decode-faster) is done in a similar way as offline decoding. Token and link structures are translated into OpenFst structures \cite{allauzen2007openfst} that present an exact lattice \cite{povey2012generating}. An exact lattice contains paths, which correspond to the candidates for ASR predictions, and stores precise costs and state-level alignments. The lattice structure enables flexible post-processing with different operations possible on lattices, such as acoustic scaling, computing the best, n-best, or oracle hypotheses, LM rescoring, lattice composition, etc. Although the lattice structure is convenient for operating with ASR output candidates before choosing the best ones, its implementation on GPUs would not be trivial. In the Kaldi GPU decoder, lattices are still created when an endpoint is reached to allow rich post-processing on CPUs~\cite{chen2018gpu}.

\subsection{Decoding on GPUs}
\label{subsec:decode_gpu}
\begin{figure*}
    \centering
    \begin{minipage}{0.35\textwidth}
    \includegraphics[width=\linewidth]{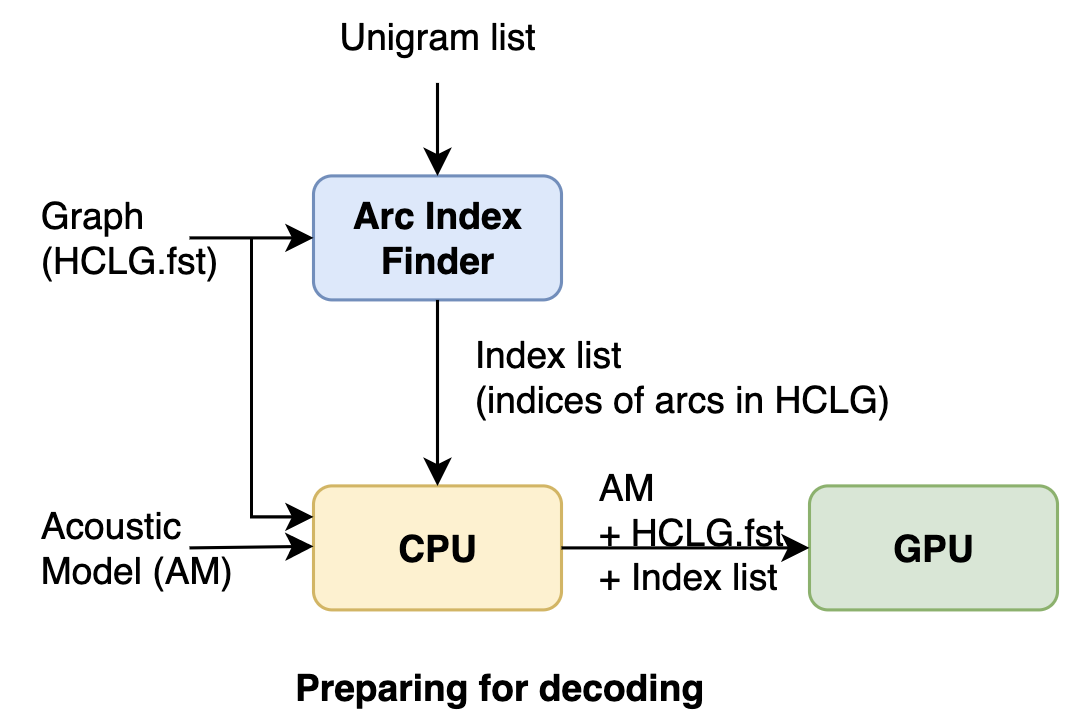}
    \end{minipage}%
    \hspace{1.15cm}
    \begin{minipage}{0.55\textwidth}
    \includegraphics[width=\linewidth]{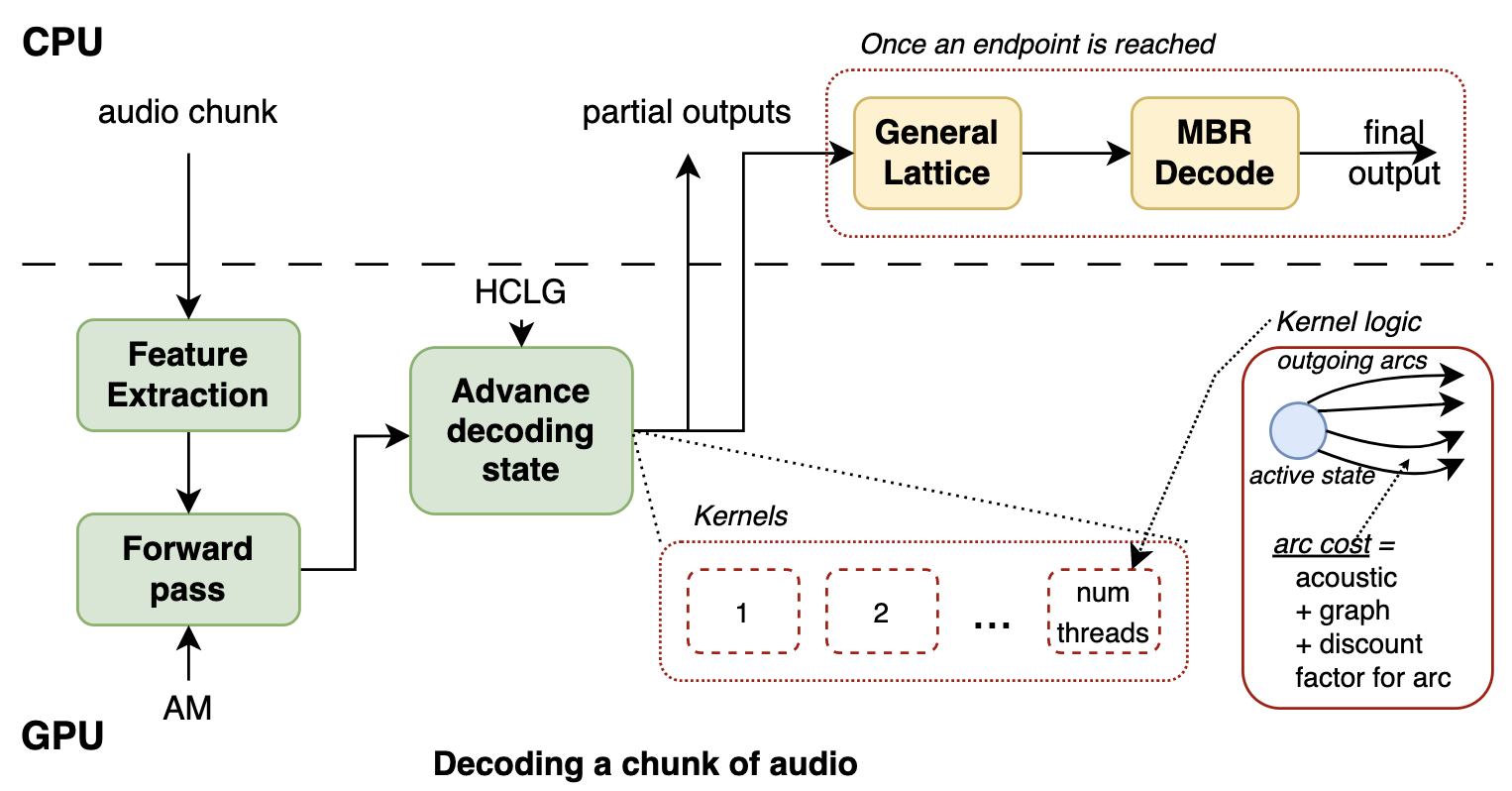}
    \end{minipage}
    \caption{Multiple stages involved in GPU decoding: initially, the models are loaded and the necessary pre-processing is done. When a chunk of audio is received, the decoding process follows with many steps split over both the CPU and the GPU.}
    \label{fig:my_label}
    \vspace{-1mm}
\end{figure*}

There are many implementations of GPU decoders~\cite{ivanov2016lvcsr,kim2014accelerating,braun2020gpu} and post-processing of their 
outputs with the CPU~\cite{kim2014accelerating,chen2018gpu,li2021parallelizable}.
For this study, we choose to work with the standard Kaldi GPU WFST decoder \cite{chen2018gpu,braun2020gpu}, as it is (1)~open-source and (2)~mostly built with Kaldi basic functions. 
The decoder\footnote{\url{https://github.com/kaldi-asr/kaldi/tree/master/src/cudadecoder}} yields up to a 240x speedup over single-core CPU decoding \cite{braun2020gpu}.
This approach has fully parallelized decoding up to the outputs, yet its current implementation does not allow any flexible rescoring. We proposed the rescoring approach inside Kaldi GPU decoder which is fully integrated into the parallelized decoding process, with no need of lattices.
%The addressed GPU decoding mechanism operates on two disparate asynchronous CUDA streams: one stream is responsible for running compute kernels, and the other one enables non-blocking device-to-host (D2H) memory copy of lattice tokens.
%Such an organization
It allows to asynchronously output intermediate results during online decoding without interrupting the computational process. The decoder pipeline first  transfers the models (i.e. acoustic model, and $HCLG$ graph) to the GPU. 
The $HCLG$ graph is represented by a special structure (\textit{cuda-fst}) on the GPU. The FST structure is represented as a set of \textit{compressed sparse rows} (CSRs) and additional metadata, which can be efficiently traversed with direct indexing \cite{braun2020gpu}.

\section{Methods for contextual biasing}
%As mentioned above, most of the rescoring methods work with lattices (or WFSTs) and composition. 
In this section, we analyze existing and the most relevant rescoring methods, choose the best implementation strategy and mention possible limitations when working on GPUs. The common feature of all methods under consideration is that contextual information is presented as a small \textit{biasing FST} built with the list of words and/or words sequences we want to boost.
%(see Fig.~\ref{fig:pic_fst}).

\subsection{Rescoring with lattice composition}
%In a hybrid ASR, system components are represented as WFSTs. Such representation allows combining different system levels with the operation of composition~\cite{mohri2002weighted}: $HCLG = H \circ C \circ L \circ G$\footnote{$H, C, L$, and $G$ are the Hidden Markov Model states, context dependency, lexicon and LM (AKA grammar) respectively.}. The operation of composition maps sequences from input transducers in a way that an output from one transducer should be an input to another one. The composition of WFSTs is convenient in use and, besides combining the basic components of ASR listed above, it can be also applied to integrate into system information from additional knowledge sources (i.e. with \textit{biasing FST}).

\textbf{Lattice rescoring.}
%Probably one of the most used ways to add contextual biasing on-the-fly is lattice rescoring when lattices from first-pass decoding are composed with a \textit{biasing FST} containing entities to boost \cite{aleksic2015bringing,hall2015composition,williams2017rescoring,serrino2019contextual,shore2012knowledge,oualil2017context,kocour21_interspeech,Iuliia_SUBMITTEDTOINTERSPEECH2021_2021}. As our goal is to avoid lattice generation, this method is not suitable for us.
One of the most used ways of contextual biasing on-the-fly is lattice rescoring in the second-pass decoding \cite{aleksic2015bringing,hall2015composition,williams2017rescoring,serrino2019contextual,shore2012knowledge,oualil2017context,kocour21_interspeech,Iuliia_SUBMITTEDTOINTERSPEECH2021_2021}. As our goal is to avoid lattice generation, this method is not suitable for us.

\noindent \textbf{HCLG ° biasing\_G.}
A rescoring approach without lattices is used in \cite{kocour21_interspeech} and assumes boosting the $HCLG$ decoding graph before decoding. The $HCLG$ graph is composed with \textit{biasing FST}, which leads to the target word weight adjustment directly in the decoding graph.
%In \cite{kocour21_interspeech}, the authors adjust weights for only single words (unigrams) in the HCLG graph, because word-strings boosting would considerably slow down the process.
If word sequences are boosted by this method, there is a limitation left: one can adjust the weights of only those sequences, which already exist in $HCLG$, yet, no new unknown sequences can be added. For example, if a 3-gram LM is used, only unigrams, bi-grams, and 3-grams already present in the LM can be boosted but not longer n-grams.

\noindent \textbf{HCL ° G\_boosted.}
The rescoring method proposed in~\cite{Iuliia_SUBMITTEDTOINTERSPEECH2021_2021} overcomes the limitation of the $HCLG \circ biasing\_G$ method. First, the $G.fst$ is separately modified in an iterative fashion in order (1)~to adjust weights of those contextual entities, which are already present in the LM, and (2)~to add new entities we want to boost, which are not present in the LM. Then, a modified $G\_boosted.fst$ is composed on-the-fly with $HCL$, allowing all necessary information from other ASR levels to be applied to all LM n-grams including newly added word sequences \cite{hori2007efficient,novak2012dynamic}.
%Although this approach seems to us to be the most optimal when speaking about $HCLG$ pre-decoding rescoring, to use it, one should have access to $HCL$ FST which is not always possible. 

\subsection{Rescoring without composition.}
In the GPU decoder, the $HCLG$ graph is represented on GPUs with the special \textit{cuda-fst} structure (see~\ref{subsec:decode_gpu}). When the graph is loaded, each outgoing arc is processed by its own thread with the \textit{load balancing expand}, generating a number of candidate tokens. The \textit{adaptive beam} is then adjusted and used to determine which candidates are added back to the main queue for further processing \cite{braun2020gpu}. GPU decoders can process multiple audio streams parallelly and it is important to enable boosting specific to an audio stream. Pre-modifying the HCLG graph in advance will result in boosting all the streams.

With a decoding graph, the rescoring approach that would be the most suitable for our goal is $HCLG \circ biasing\_G$ composition. Instead of composition, weights in the $HCLG$ could be adjusted iteratively, like for \textit{G} in the $HCL \circ G\_boosted$ method, or by their indices that would allow flexible and less computationally expensive rescoring. As decoding on GPUs we cannot afford composition with $HCL$ following the $G$ rescoring, it is not possible to add unknown word sequences to the graph. Thus, the n-gram set that we can boost is always limited by the LM, like in the $HCLG \circ biasing\_G$ method. The contextual \textit{boosting} information for this method can be passed (1)~as a \textit{biasing} FST, or (2)~as a list of entities we want to boost, where all words are replaced with their IDs from the symbol table.
%(Fig.~\ref{fig:pic_fst}))

\section{Rescoring in GPU decoder}
%We use a WFST decoder, which is capable of online streaming, as well as offline batch processing of audio, using GPUs \cite{chen2018gpu,braun2020gpu}.
In our implementation of rescoring, we focus on three tasks: (1)~unigrams boosting, (2)~word sequence boosting, (3)~dynamic update of contextual information, i.e. biasing FST. Another important aspect is to make sure that our implementation is optimal and that the decoding slowdown is minimized.

\subsection{Implementation}
\label{subsec:boosting_uni}

The $HCLG$ graph is represented as a set of \textit{compressed sparse rows} (CSRs) and additional metadata stored in memory. Before CSRs are generated, the arc information from the loaded $HCLG$ graph is temporarily kept in separate vectors: for input labels IDs, output labels IDs, next state IDs, and weights. This information is further copied to the GPUs. In order to rescore on GPUs, along with the decoding FST, we
%additionally
load the biasing FST, whose arc information is also saved in vectors but only for those arcs that should be boosted. Algorithm~\ref{alg:pseudocode} gives the pseudocode of the procedure to determine which arcs to boost given a list of words. 
The algorithm is an extension of Depth First Search to find a sequence of arcs that would generate the desired sequence.
We also pay attention to the possibility that the first word in the sequence may
%actually
start in the middle of the utterance.
During decoding, if an arc index coincides with any token index saved from the current biasing FST, the arc's weight is adjusted by the discount factor. The boosting weight is the sum of the original arc's weight and the discount factor. The discount factor we use equals -2.0 which was empirically identified in the previous studies \cite{Iuliia_SUBMITTEDTOINTERSPEECH2021_2021}.  
The process is illustrated in Figure~\ref{fig:my_label}.

\setlength{\textfloatsep}{0pt}
\begin{algorithm}[t!]
\DontPrintSemicolon
\SetKwInOut{KwIn}{Input}
\SetKwInOut{KwOut}{Output}
\SetKwData{Fst}{fst}
\KwIn{\Fst: decoding graph; $w_1 w_2 ... w_k$: word sequence to boost, $N$: N value of N-gram language model}
\KwOut{arcs\_indices: arcs that need to be boosted}
arcs\_indices = Set()\;
statesReached = GetStatesThatOutputToken(fst, $w_1$)\;
\For{$t\leftarrow 2$ \KwTo $k$} {
    prevStatesReached = statesReached\;
    statesReached = set()\;
    \For{$s_p$ $\leftarrow$ prevStatesReached} {
        \textit{$//$ DepthFirstSearchSpecial takes first edge with $w_t$ and then considers only edges that output $\epsilon$ until $w_t$ is output}\\
        reachableStates = DepthFirstSearchSpecial(fst, $s_p$, $w_t$)\;
        \textit{$//$ now we know we can emit the next token}\\
        statesReached.add(reachableStates)\;
        arcs\_indices.Add(ArcsIndicesWithOutput($s_p$, $w_{t-1}$))\;
    }
}
\For{$s$ $\leftarrow$ statesReached} {
    arcs\_indices.Add(ArcsIndicesWithOutput($s$, $w_{t}$))\;
}
\Return{arcs\_indices}\;
\vspace{1mm}
\caption{Pseudocode to find the arcs to be boosted given a word sequence $(w_{1} w_{2} ... w_{k})$ \label{alg:pseudocode}}
    % \For{$arc\leftarrow 1$ \KwTo aiter(FST, currToken)} {
    %     \If{currToken == $arc$.olabel} {
    %         unique\_arc\_indices.Add($arc$)
    %     }}
% TODO: fix formatting
\end{algorithm}

\begin{table*}[t]
  \caption{Contextual biasing with online CPU and GPU decoders (GT is a ground truth sequence (available only for ATCO2 sets); `partial hypotheses' are real-time model predictions; EntWER is a WER calculated for biased entities only).}
  \label{tab:results_gpu_boosting}
  \centering
      \begin{tabular}{ lccccccc }
    \toprule
& \multicolumn{2}{c}{\textbf{ATCO2-4h}} & \multicolumn{2}{c}{\textbf{ATCO2-public}} & \multicolumn{3}{c}{\textbf{Earnings21}} \\
    & \textbf{WER} & \textbf{EntWER} & \textbf{WER} & \textbf{EntWER} & \textbf{WER} & \textbf{EntWER} & \textbf{RTFX} \\

    \midrule
    \multicolumn{2}{c}{\textbf{Online decoding on CPU}} & & & & & & \\
    \midrule
    \textbf{No biasing} & 32.6 & 36.4 & 24.3 & 26.4 & 21.6 & 59.0 & 7.001 \\
    \textbf{Biased unigrams (partial hypotheses)} & 34.6 & 35.4 & 25.0 & 25.7 & - & - & - \\
    \textbf{Biased sequences (partial hypotheses)} & 32.5 & 34.3 & 24.0 & 24.2 & 21.7 & 51.8 & 3.577 \\
    \textbf{Biased GT (partial hypotheses)} & 31.0 & 30.4 & 23.1 & 20.3 & - & - & - \\
    \midrule
    \multicolumn{2}{c}{\textbf{Online decoding on GPU}} & & & & & & \\
    \midrule
    \textbf{No biasing} & 32.2 & 36.3 & 24.5 & 26.4 & 21.4 & 60.5 & 26.062  \\
    \textbf{Biased unigrams (at endpoints)} & 34.1 & 35.7 & 25.0 & 25.4 & - & - & - \\
    \textbf{Biased sequences (at endpoints)} & 31.2 & 34.4 & 24.0 & 24.1 & 21.4 & 52.4 & 26.061 \\
    \textbf{Biased GT (at endpoints)} & 30.5 & 30.1 & 23.4 & 21.2 & - & - & - \\
    \midrule
    \textbf{Biased unigrams (partial hypotheses)} & 33.2 & 35.5 & 24.7 & 25.1 & - & - & - \\
    \textbf{Biased sequences (partial hypotheses)} & 32.9 & 34.6 & 24.9 & 25.3 & 22.2 & 52.7 & 26.065 \\
    \textbf{Biased GT (partial hypotheses)} & 30.7 & 29.4 & 23.8 & 21.9 & - & - & - \\

    \bottomrule
  \end{tabular}
\end{table*}

\begin{table}[t]
  \caption{Test sets with context information (number of biasing entities for ATCO2 sets$^\dagger$ is given on average per utterance).}
  \label{tab:test_sets}
  \centering
  \scalebox{0.9}{
  \begin{tabular}{ lccc }
    \toprule
    \textbf{Test set} & \textbf{Size} & \textbf{Hours} & \textbf{Biasing entities} \\
    \midrule
    ATCO2 & 3535 utterances & 4 & 214$^\dagger$ \\
    ATCO2-public & 871 utterances & 1 & 140$^\dagger$ \\
    Earnings21 & 44 interviews & 39 & 1013 \\
    \bottomrule
  \end{tabular}}
\end{table}

\subsection{Boosting unigrams and word sequences}
%As mentioned in the previous subsection, 
For every contextual biasing FST, we keep track of indices of the arcs to boost, which are identified by Algorithm \ref{alg:pseudocode}.
In the decoding kernels, this adds an extra cost of searching if the current thread is processing an arc to be boosted.
To enable faster search, we store the indices sorted, and perform a simple binary search.
An additional complexity of $O(\log k)$ is added to each processing thread.
This is negligible since $k$ is significantly less than the total number of arcs in the graph.
Compared to excessive memory requirements if storing separate decoding graphs for each context, we only store few 100s of integers. 

The size of biasing FST depends on the number of entities to boost. As with the increasing number of contextual entities, the biasing effect typically goes down\footnote{The optimal size of biasing FST highly depends on the data; in \cite{chen2019end}, the performance began to degrade when a number of contextual entities exceeded 1000.}, we assume that the size of biasing FST stays small not to exceed available memory. In our experiments, the largest FST has 1013 entities (Table~\ref{tab:test_sets}), and the number of \textit{boosting} arcs we keep in memory per FST is significantly less than the total number of arcs since we aim to boost only the arcs related to the entities we are interested in. 

\subsection{Dynamic context update}
To provide flexible biasing when the context is modified, we introduce the functionality of a dynamic switch between different biasing FSTs. We assume that certain context situations are anticipated in advance and the corresponding biasing FSTs are available before decoding starts. All expected biasing FSTs are pre-loaded and saved in separate vectors similar to how it is described in \ref{subsec:boosting_uni}. Depending on the context a needed \textit{biasing FST} indices are used to adjust the corresponding arc weights.

\section{Data and experimental setup}
\subsection{Data}
For biasing experiments we need test sets which along with text transcriptions would also have biasing list(s) with entities to boost. There are only few publicly available test sets that satisfy this criterion. One of the test sets we use is publicly available \textit{Earning21} \cite{delrio2021earnings21} which has been recently updated with two biasing lists based on the NER\footnote{The two biasing lists are the \textit{oracle} and the \textit{distractor} lists: \url{https://github.com/revdotcom/speech-datasets/tree/main/earnings21/bias_lists}. For our experiments, we use only the \textit{oracle} list.} \cite{drexler2022improving}. The Earnings21 
biasing lists contain both unigrams and word sequences, and we keep them together
%\footnote{In Table~\ref{tab:results_gpu_boosting}, it is categorized as sequence boosting.}.
(in Table~\ref{tab:results_gpu_boosting}, it is categorized as sequence boosting).
For decoding, we split the audios into 3-minute long segments, similar to~\cite{drexler2022improving}.
In order to test the proposed algorithm for the case when the biasing context is always changing, we additionally use two test sets from the ATC domain. ATC conversations are usually saturated with \textit{callsigns}\footnote{Callsigns are unique identifiers for air crafts, of which the first part is an abbreviation of the airline name and the last part is a flight number that contains a digit combination and may also incorporate an additional character combination, e.g., \textit{ryanair one sierra golf}.} used to address air crafts, and \textit{contextual data} is constantly coming from the radar that registers those callsigns of corresponding air crafts that are currently in the airspace \cite{kocour21_interspeech,Iuliia_SUBMITTEDTOINTERSPEECH2021_2021,nigmatulina2022two}. One ATC test set is ATCO2~\cite{zuluaga2022atco2,zuluaga2023lessons} and another one is a publicly available 1-hour long subset of ATCO2, referred to as ATCO2-1h~\cite{zuluaga2023does}.\footnote{Website: \url{https://www.atco2.org/data}} Each utterance in the ATC sets is provided with a list of callsigns to bias and with the ground truth callsign, or NO\_CALLSIGN if an utterance does not contain any. ATCO2 biasing lists contain word sequences, and for the unigram boosting we converted them into lists of unique single words. All biasing lists include about 10\% of OOV words. All used sets are English data; an overview including the number of biasing entities is given in Table~\ref{tab:test_sets}.
%{\color{red}add a publicly available test set.}
%\footnote{\url{http://catalog.elra.info/en-us/repository/browse/ELRA-S0484/}}

%\begin{table}[t]
%  \caption{Results of offline boosting on liveATC test set (EntWER is a WER on boosted entities only)}
%  \label{tab:results_gpu_boosting}
%  \centering
%  \begin{tabular}{ llccc }
%    \toprule
%    \multicolumn{2}{c}{\textbf{Decoding type}} & \textbf{WER} & \textbf{EntWER} & \textbf{RTFX} \\
    %\cline{3-5}
    %& & & \textit{unigrams} & \textit{word sequences} \\
%    \midrule
%    CPU & baseline & 28.3 & 29.7 & 39.2 \\
%    \cline{2-5}
%     & boosted & 27.5 & 25.6 & 39.3 \\
%    \midrule
%    GPU & baseline & 29.1 & 28.8 & 373.6 \\
%    \cline{2-5}
%     & boosted & 28.8 & 26.5 & 439.7 \\
%    \bottomrule
%  \end{tabular}
%\end{table}

\subsection{ASR model}
\label{subsec:model}
For the acoustic models, we use the Kaldi toolkit~\cite{povey2011kaldi}. For the experiments on ATC data, we trained a CNN-TDNNF model with $\approx$1200 hours of ATC labeled data after 3-fold speed perturbation. The system follows the standard Kaldi recipe with MFCC and i-vectors features; the standard chain training is based on LF-MMI~\cite{povey2016purely,madikeri2020lattice} including one-third frame sub-sampling. The LM is 3-gram model trained on the same data as the acoustic model with additional text data coming from public resources such as airlines names, airports, the ICAO alphabet, and way-points in Europe. For the experiments on the Earnings21 set, we use the pre-trained chain LSTM-TDNN Kaldi model\footnote{\url{https://kaldi-asr.org/models/m14}} with Gigaspeech-XL speech corpus \cite{chen2021gigaspeech}.

\subsection{Evaluation of speed}
To demonstrate the lack of difference in the decoding time with VS without biasing, we measured relative decoding time with the \textit{inverse real-time factor (RTFX)}, which is the ratio between the length of the processed audio and the decoding time. The RTFX is measured with 1 GPU NVIDIA GeForce RTX 3090, with 2K clients, and on 81 minutes of Earnings21 data, which are split into 27 utterances, each of 3 minutes in length.

\section{Results}
We compare WER results achieved (1)~with contextual biasing on CPUs VS GPUs with lattice rescoring at \textit{endpoints} VS dynamic biasing for \textit{partial hypotheses} on GPUs (to see if there is performance degradation when rescoring is done without composition), (2)~on GPUs with VS without applying contextual biasing (to see how the method improves the recognition). We do not compare the performance of our implementation to previous work, as there is no such results for ATCO2 sets, and biasing results on Earnings21 \cite{drexler2022improving} are achieved with an End-to-End model with a different biasing approach which would be incomparable to our experiments. The results of the partial hypotheses biasing on GPUs are taken directly from the final server outputs, i.e. before it is sent for post-processing. Table~\ref{tab:results_gpu_boosting} reports the results with utterance WER, and entities WER (EntWER) where WER is calculated only on biased entities.
Comparing the performance of biasing with CPU decoder to the one on GPUs, the achieved improvement is almost the same, when rescored with lattices.
Contextual biasing on GPUs always helps improve performance on the entities: e.g. EntWER 52.4\% instead of 60.5\%, resulting in a relative improvement of 13.4\% on Earnings21.
The results in utterance WERs stay the same or slightly improve when sequences are boosted. 
Dynamic biasing of partial hypotheses on GPUs slightly differs from the other results, as weights are modified directly in the HCLG graph instead of decoder output candidates. Overall, the performance of dynamic biasing on GPUs shows similar improvement on entities over the baseline compared to the lattice composition approach.

The main advantages of our implementation are its speed and flexibility. Decoding on GPUs allows a considerable increase in speed compared to CPUs. Adding boosting inside the GPU decoder does not slow down the decoding process with the RTFX staying almost the same: 26.06. Pre-biasing the HCLG graph in advance would lead to similar improvements but does not allow dynamic context adaptation. The main limitation is that it is not possible to add unknown word sequences to the graph and the n-gram set we can boost is always limited by the LM.
%In the future, the method can be further improved to enable RNN-LM rescoring.
%; in this study, we did not experiment with RNN-LM rescoring, as we stayed focused on the dynamic context updated and RNN-LM would be too big for this case.
In the future, the method can be also extended to WFST decoding for End-to-End models, where instead of words End-to-End model units, i.e. characters or subwords, are used.
%The method can be also extended to biasing word sequences that are present in the decoding graph.
%As in the implementation we keep all the information of arcs to be boosted, it can be the next step of the improvement.

%Additionally, we compare performance when biasing is done on unigrams vs word sequences. As target context entities in our experiments are word sequences, word sequence boosting has a stronger effect than unigram boosting, although unigram boosting still improved over the baseline.

%boosting sequences that are already present in the decoding graph.

% TODO: do we need to add speed?

%\begin{table}[t]
%  \caption{Examples of improved callsign recognition (bold part)}
%  \label{tab:improve_examples}
%  \centering
%  \resizebox{0.47\textwidth}{!}{
%  \begin{tabular}{ p{3.7cm} p{4.cm} }
%    \toprule
%    Baseline (incorrect callsign) & Boosted (correct callsign) \\
%    \midrule
%    \textbf{wizz air} four one six & \textbf{iceair} four one six \\
%    \textbf{easy} three delta & \textbf{fraction eight eight} three delta \\
%    \textbf{serbia} one nine lima & \textbf{stobart} one nine lima \\
%    \bottomrule
%  \end{tabular}
%  }
%\end{table}

\section{Conclusion}
Motivated by the high effectiveness of contextual biasing on CPUs, we proposed an algorithm and its implementation for dynamic contextual biasing on GPUs for real-time hypotheses. Given the context words and word sequences as input the method adjusts target arc weights in the decoding graph in a distributed way and without lattice generation. This approach allows fast and flexible adaptation to a current context and is a step toward closer integration between inference and decoding.
A relative improvement in EntityWER of 13.4\% was achieved on the Earnings21 set when biasing the target entities.

\bibliographystyle{IEEEtran}
\bibliography{references}

\begin{thebibliography}{10}
\providecommand{\url}[1]{#1}
\csname url@samestyle\endcsname
\providecommand{\newblock}{\relax}
\providecommand{\bibinfo}[2]{#2}
\providecommand{\BIBentrySTDinterwordspacing}{\spaceskip=0pt\relax}
\providecommand{\BIBentryALTinterwordstretchfactor}{4}
\providecommand{\BIBentryALTinterwordspacing}{\spaceskip=\fontdimen2\font plus
\BIBentryALTinterwordstretchfactor\fontdimen3\font minus
  \fontdimen4\font\relax}
\providecommand{\BIBforeignlanguage}[2]{{%
\expandafter\ifx\csname l@#1\endcsname\relax
\typeout{** WARNING: IEEEtran.bst: No hyphenation pattern has been}%
\typeout{** loaded for the language `#1'. Using the pattern for}%
\typeout{** the default language instead.}%
\else
\language=\csname l@#1\endcsname
\fi
#2}}
\providecommand{\BIBdecl}{\relax}
\BIBdecl

\bibitem{aleksic2015bringing}
P.~Aleksic, M.~Ghodsi, A.~Michaely, C.~Allauzen, K.~Hall, B.~Roark, D.~Rybach,
  and P.~Moreno, ``Bringing contextual information to google speech
  recognition,'' in \emph{Proc. of Interspeech}, 2015, pp. 468--472.

\bibitem{hall2015composition}
K.~Hall, E.~Cho, C.~Allauzen, F.~Beaufays, N.~Coccaro, K.~Nakajima, M.~Riley,
  B.~Roark, D.~Rybach, and L.~Zhang, ``Composition-based on-the-fly rescoring
  for salient n-gram biasing,'' in \emph{Proc. of Interspeech}, 2015, pp.
  1418--1422.

\bibitem{williams2017rescoring}
I.~Williams and P.~S. Aleksic, ``Rescoring-aware beam search for reduced search
  errors in contextual automatic speech recognition.'' in \emph{Proc. of
  Interspeech}, 2017, pp. 508--512.

\bibitem{serrino2019contextual}
J.~Serrino, L.~Velikovich, P.~S. Aleksic, and C.~Allauzen, ``Contextual
  recovery of out-of-lattice named entities in automatic speech recognition.''
  in \emph{Interspeech}, 2019, pp. 3830--3834.

\bibitem{shore2012knowledge}
T.~Shore, F.~Faubel, H.~Helmke, and D.~Klakow, ``Knowledge-based word lattice
  rescoring in a dynamic context,'' in \emph{Proc. of Interspeech}, 2012, pp.
  1083--1086.

\bibitem{oualil2017context}
Y.~Oualil, D.~Klakow, G.~Szasz{\'a}k, A.~Srinivasamurthy, H.~Helmke, and
  P.~Motlicek, ``A context-aware speech recognition and understanding system
  for air traffic control domain,'' in \emph{2017 IEEE Automatic Speech
  Recognition and Understanding Workshop (ASRU)}.\hskip 1em plus 0.5em minus
  0.4em\relax IEEE, 2017, pp. 404--408.

\bibitem{kocour21_interspeech}
M.~Kocour, K.~Vesel{\`y}, A.~Blatt, J.~Z. Gomez, I.~Sz{\"o}ke, J.~Cernocky,
  D.~Klakow, and P.~Motlicek, ``{Boosting of Contextual Information in ASR for
  Air-Traffic Call-Sign Recognition},'' in \emph{Proc. Interspeech 2021}, 2021,
  pp. 3301--3305.

\bibitem{Iuliia_SUBMITTEDTOINTERSPEECH2021_2021}
I.~Nigmatulina, R.~Braun, J.~Zuluaga-Gomez, and P.~Motlicek, ``Improving
  callsign recognition with air-surveillance data in air-traffic
  communication,'' Idiap Research Institute.\hskip 1em plus 0.5em minus
  0.4em\relax Idiap Research Institute, 2021, pp. 1--5.

\bibitem{zuluagagomez21_interspeech}
J.~Zuluaga-Gomez, I.~Nigmatulina, A.~Prasad, P.~Motlicek, K.~Vesel{\`y},
  M.~Kocour, and I.~Sz{\"o}ke, ``{Contextual Semi-Supervised Learning: An
  Approach to Leverage Air-Surveillance and Untranscribed \MakeUppercase{ATC}
  Data in \MakeUppercase{ASR} Systems},'' in \emph{Interspeech}, 2021, pp.
  3296--3300.

\bibitem{motlicek2010english}
P.~Motlicek, F.~Valente, and P.~N. Garner, ``English spoken term detection in
  multilingual recordings,'' in \emph{Eleventh Annual Conference of the
  International Speech Communication Association}, 2010.

\bibitem{braun2020gpu}
H.~Braun, J.~Luitjens, R.~Leary, T.~Kaldewey, and D.~Povey, ``Gpu-accelerated
  viterbi exact lattice decoder for batched online and offline speech
  recognition,'' in \emph{ICASSP 2020-2020 IEEE International Conference on
  Acoustics, Speech and Signal Processing (ICASSP)}.\hskip 1em plus 0.5em minus
  0.4em\relax IEEE, 2020, pp. 7874--7878.

\bibitem{povey2011kaldi}
D.~Povey, A.~Ghoshal, G.~Boulianne, L.~Burget, O.~Glembek, N.~Goel,
  M.~Hannemann, P.~Motlicek, Y.~Qian, P.~Schwarz \emph{et~al.}, ``The kaldi
  speech recognition toolkit,'' in \emph{IEEE workshop on automatic speech
  recognition and understanding}, no. CONF.\hskip 1em plus 0.5em minus
  0.4em\relax IEEE Signal Processing Society, 2011.

\bibitem{argueta2018composing}
A.~Argueta and D.~Chiang, ``Composing finite state transducers on gpus,'' in
  \emph{Proc. of the 56th Annual Meeting of the Association for Computational
  Linguistics (Vol. 1)}, 2018, pp. 2697--2705.

\bibitem{li2021parallelizable}
K.~Li, D.~Povey, and S.~Khudanpur, ``A parallelizable lattice rescoring
  strategy with neural language models,'' in \emph{ICASSP 2021-2021 IEEE
  International Conference on Acoustics, Speech and Signal Processing
  (ICASSP)}.\hskip 1em plus 0.5em minus 0.4em\relax IEEE, 2021, pp. 6518--6522.

\bibitem{nigmatulina2022two}
I.~Nigmatulina, J.~Zuluaga-Gomez, A.~Prasad, S.~S. Sarfjoo, and P.~Motlicek,
  ``A two-step approach to leverage contextual data: speech recognition in
  air-traffic communications,'' in \emph{ICASSP 2022-2022 IEEE International
  Conference on Acoustics, Speech and Signal Processing (ICASSP)}.\hskip 1em
  plus 0.5em minus 0.4em\relax IEEE, 2022, pp. 6282--6286.

\bibitem{allauzen2007openfst}
C.~Allauzen, M.~Riley, J.~Schalkwyk, W.~Skut, and M.~Mohri, ``Openfst: A
  general and efficient weighted finite-state transducer library,'' in
  \emph{International Conference on Implementation and Application of
  Automata}.\hskip 1em plus 0.5em minus 0.4em\relax Springer, 2007, pp. 11--23.

\bibitem{povey2012generating}
D.~Povey, M.~Hannemann, G.~Boulianne, L.~Burget, A.~Ghoshal, M.~Janda,
  M.~Karafi{\'a}t, S.~Kombrink, P.~Motl{\'\i}{\v{c}}ek, Y.~Qian \emph{et~al.},
  ``Generating exact lattices in the wfst framework,'' in \emph{2012 IEEE
  International Conference on Acoustics, Speech and Signal Processing
  (ICASSP)}.\hskip 1em plus 0.5em minus 0.4em\relax IEEE, 2012, pp. 4213--4216.

\bibitem{chen2018gpu}
Z.~Chen, J.~Luitjens, H.~Xu, Y.~Wang, D.~Povey, and S.~Khudanpur, ``A gpu-based
  wfst decoder with exact lattice generation,'' in \emph{Proc. of Interspeech},
  2018, pp. 2212--2216.

\bibitem{ivanov2016lvcsr}
A.~V. Ivanov, P.~L. Lange, and D.~Suendermann-Oeft, ``Lvcsr system on a hybrid
  gpu-cpu embedded platform for real-time dialog applications,'' in
  \emph{Proceedings of the 17th Annual Meeting of the Special Interest Group on
  Discourse and Dialogue}, 2016, pp. 220--223.

\bibitem{kim2014accelerating}
J.~Kim and I.~Lane, ``Accelerating large vocabulary continuous speech
  recognition on heterogeneous cpu-gpu platforms,'' in \emph{Proc. of IEEE
  International Conference on Acoustics, Speech and Signal Processing
  (ICASSP)}.\hskip 1em plus 0.5em minus 0.4em\relax IEEE, 2014, pp. 3291--3295.

\bibitem{hori2007efficient}
T.~Hori, C.~Hori, Y.~Minami, and A.~Nakamura, ``Efficient wfst-based one-pass
  decoding with on-the-fly hypothesis rescoring in extremely large vocabulary
  continuous speech recognition,'' \emph{IEEE Transactions on audio, speech,
  and language processing}, vol.~15, no.~4, pp. 1352--1365, 2007.

\bibitem{novak2012dynamic}
J.~R. Novak, N.~Minematsu, and K.~Hirose, ``Dynamic grammars with lookahead
  composition for wfst-based speech recognition,'' in \emph{Thirteenth Annual
  Conference of the International Speech Communication Association}, 2012.

\bibitem{chen2019end}
Z.~Chen, M.~Jain, Y.~Wang, M.~L. Seltzer, and C.~Fuegen, ``End-to-end
  contextual speech recognition using class language models and a token passing
  decoder,'' in \emph{Proc. of IEEE International Conference on Acoustics,
  Speech and Signal Processing (ICASSP)}.\hskip 1em plus 0.5em minus
  0.4em\relax IEEE, 2019, pp. 6186--6190.

\bibitem{delrio2021earnings21}
M.~D. Rio, N.~Delworth, R.~Westerman, M.~Huang, N.~Bhandari, J.~Palakapilly,
  Q.~McNamara, J.~Dong, P.~Zelasko, and M.~Jette, ``Earnings-21: A practical
  benchmark for asr in the wild,'' in \emph{Interspeech}, 2021.

\bibitem{drexler2022improving}
J.~Drexler~Fox and N.~Delworth, ``Improving contextual recognition of rare
  words with an alternate spelling prediction model,'' in \emph{Interspeech},
  2022, pp. 3914--3918.

\bibitem{zuluaga2022atco2}
J.~Zuluaga-Gomez, K.~Vesel{\`y}, I.~Sz{\"o}ke, P.~Motlicek, M.~Kocour,
  M.~Rigault, K.~Choukri, A.~Prasad, S.~S. Sarfjoo, I.~Nigmatulina
  \emph{et~al.}, ``{ATCO2 corpus: A Large-Scale Dataset for Research on
  Automatic Speech Recognition and Natural Language Understanding of Air
  Traffic Control Communications},'' \emph{arXiv preprint arXiv:2211.04054},
  2022.

\bibitem{zuluaga2023lessons}
J.~Zuluaga-Gomez, I.~Nigmatulina, A.~Prasad, P.~Motlicek, D.~Khalil,
  S.~Madikeri, A.~Tart, I.~Szoke, V.~Lenders, M.~Rigault \emph{et~al.},
  ``{Lessons Learned in ATCO2: 5000 hours of Air Traffic Control Communications
  for Robust Automatic Speech Recognition and Understanding},'' \emph{arXiv
  preprint arXiv:2305.01155}, 2023.

\bibitem{zuluaga2023does}
J.~Zuluaga-Gomez, A.~Prasad, I.~Nigmatulina, S.~S. Sarfjoo, P.~Motlicek,
  M.~Kleinert, H.~Helmke, O.~Ohneiser, and Q.~Zhan, ``How does pre-trained
  wav2vec 2.0 perform on domain-shifted asr? an extensive benchmark on air
  traffic control communications,'' in \emph{2022 IEEE Spoken Language
  Technology Workshop (SLT)}.\hskip 1em plus 0.5em minus 0.4em\relax IEEE,
  2023, pp. 205--212.

\bibitem{povey2016purely}
D.~Povey, V.~Peddinti, D.~Galvez, P.~Ghahremani, V.~Manohar, X.~Na, Y.~Wang,
  and S.~Khudanpur, ``Purely sequence-trained neural networks for asr based on
  lattice-free mmi.'' in \emph{Interspeech}, 2016, pp. 2751--2755.

\bibitem{madikeri2020lattice}
S.~R. Madikeri, B.~K. Khonglah, S.~Tong, P.~Motlicek, H.~Bourlard, and
  D.~Povey, ``Lattice-free maximum mutual information training of multilingual
  speech recognition systems.'' in \emph{Interspeech}, 2020, pp. 4746--4750.

\bibitem{chen2021gigaspeech}
G.~Chen, S.~Chai, G.~Wang, J.~Du, W.-Q. Zhang, C.~Weng, D.~Su, D.~Povey,
  J.~Trmal, J.~Zhang \emph{et~al.}, ``Gigaspeech: An evolving, multi-domain asr
  corpus with 10,000 hours of transcribed audio,'' in \emph{Interspeech}, 2021.

\end{thebibliography}

\end{document}